\documentclass[10pt,draft]{dis03}
\usepackage{amsmath,epsfig}

\begin{document}
\title{BEAUTY QUARK PRODUCTION IN DEEP INELASTIC SCATTERING AT HERA \\
\vspace*{-35mm}{\it \begin{flushleft} \small
Talk presented at the XI International Workshop on Deep Inelastic Scattering, \\
23-27 April 2003, St.~Petersburg, Russia.
\end{flushleft}
}\vspace*{21.6mm}
}

\author{Vincenzo Chiochia\footnote{Now at the Physik Institut der Universit\"at Z\"urich-Irchel, 
8057 Z\"urich, Switzerland.}\\
DESY, Deutsches Elektronen Synchrotron \\
22607 Hamburg, Germany\\
E-mail: vincenzo.chiochia@cern.ch}

\maketitle
%
%
\begin{abstract}
\noindent The measurement of the beauty quark production cross section in deep 
inelastic scattering with the ZEUS detector at HERA is presented. 
The cross section for beauty quarks, measured in the kinematic range $Q^2 > 2$ GeV$^2$ 
and $0.05 < y < 0.7$, and for events with at least one muon and one jet in the Breit frame,
is somewhat above the next-to-leading order (NLO) QCD prediction, but in agreement
within the experimental and theoretical uncertainties.
\end{abstract}
%
%
%
\section{Introduction}

Deep inelastic scattering (DIS) offers the unique opportunity to study the
production mechanism of heavy quarks via the strong interaction in a particularly
clean environment where a point-like projectile, a virtual photon with a virtuality
$Q^2$, collides with a proton.
A first measurement of the total inclusive $b$-quark cross section in the DIS region has 
been released by the H1 collaboration \cite{H1:1013}.
Recently, the ZEUS collaboration has measured the production cross section of $b$-quarks in DIS at HERA,
in the reaction with at least one hard jet in the Breit frame\footnote{In the Breit frame, 
defined by $\vec{\gamma}+2x\vec{P} = \vec{0}$, where $\vec{\gamma}$ is the momentum of the
exchanged photon, $x$ is the Bjorken scaling variable and $\vec{P}$ is the
proton momentum, a purely space-like photon and a proton collide head-on.} 
and a muon in the final state ($ep \rightarrow e b \bar{b} X \rightarrow e~\mbox{jet}~\mu~X$)
\cite{ZEUS:783}.

Due to the large $b$-quark mass, muons from semi-leptonic $b$ decays usually
have high values of $p_T^{rel}$, which is the transverse momentum of the muon
with respect to the axis of the closest jet. This allows a statistical separation
of the signal from the background.

\section{Event selection and extraction of the b decays}

The data used in the measurement were collected during the 1999-2000 HERA
running period, corresponding to a total integrated luminosity of 60
pb$^{-1}$ and at a center-of-mass energy of 318 GeV. A detailed description
of the ZEUS detector can be found elsewhere \cite{Derrick:1992kk}.

The event sample was selected in four steps:
\begin{itemize}
\item DIS events were selected by requiring a well reconstructed outgoing 
  po\-si\-tron with energy greater than 10 GeV, $Q^2 > 2$ GeV$^2$ and inelasticity
  $0.05<y<0.7$, where $y=Q^2/xs$. 
\item Muons were identified by requiring a reconstructed segment both in the
  inner and outer part of barrel and rear muon chambers, where the rear direction is
  defined by the in\-co\-ming po\-si\-tron. The reconstructed muons are matched in space
  and momentum with a track found in the central tracker. In addition, a cut on the muon momentum 
  $p^\mu > 2$ GeV is applied together with a cut on the muon polar angle $\theta^\mu$,
  $30^\circ < \theta^\mu < 160^\circ$.
\item Hadronic final-state objects, reconstructed from tracks and energy
  deposit in the calorimeter, were boosted to the Breit frame and clustered into 
  jets using the $k_T$ clustering algorithm \cite{Catani:1993hr}. Selected events had at least one
  jet with $E_T^{Breit} > 6$ GeV and within the detector acceptance $-2 < \eta^{Lab} < 2.5$,
  where $\eta^{Lab}$ is the pseudorapidity in the laboratory frame\footnote{The pseudorapidity
    $\eta$ is defined as $\eta = -\ln(\tan\theta/2)$, where $\theta$ is the polar angle measured
    in with respect to the proton beam direction}.
\item The muons in the sample were associated with a jet using the $k_T$ 
  algorithm information, where the associated jet was not necessarily the jet
  satisfying the cuts above. To ensure a good reconstruction of the associated
  jet, it was required to have transverse energy in the Breit frame $E_T^{Breit} > 4$ GeV.
\end{itemize}
After all selection cuts the final data sample contains 836 events.

To correct the results for detector effects and to extract the fraction
of events from $b$ decays the {\tt RAPGAP} Monte Carlo program \cite{Jung:1993gf} was used.
The light-flavour and charm-quark {\tt RAPGAP} samples were mixed according to the 
relative luminosities and the $b$-quark sample was added according to the 
beauty fraction determined from the analysis of the $p_T^{rel}$ distribution.
To determine the fraction of events from $b$ decays in the data, the contribution
from light-plus-charm flavours and beauty were allowed to vary, and the
best mixture was extracted using a binned maximum-likelihood method. The measured 
fraction of $b$ decays is $(25\pm5)\%$, where the error is statistical.
This mixture describes well the shape of all event kinematic variables.

\section{Results}

The total visible cross section, $\sigma_{b\bar{b}}$, was determined in the 
kinematic range $Q^2 > 2$ GeV$^2$, $0.05<y<0.7$ with a muon with $p^\mu>2$ GeV
and $30^\circ < \theta^\mu < 160^\circ$ and at least one jet in the Breit frame with
$E_T^{Breit} > 6$ GeV and $-2 < \eta^{Lab} < 2.5$. The cross section is
\begin{equation}
\sigma_{b\bar{b}}(e^+p \rightarrow e^+b\bar{b}X \rightarrow e^+~\mbox{jet}~\mu~ X)
= 38.7 \pm 7.7~\mbox{(stat.)}^{+6.1}_{-5.0}\mbox{(syst.)}~\mbox{pb}.
\end{equation}
The NLO QCD predictions were evaluated using the {\tt HVQDIS} program 
\cite{Harris:1997zq}.
Fragmentation of $b$-quarks into hadrons was performed using the Peterson
function \cite{Peterson:1982ak} with $\epsilon=0.002$ as suggested in Ref.~\cite{Cacciari:2002pa}. 
The semi-leptonic decay of $b$-hadrons was modeled using a parameterisation 
of the muon momentum spectrum extracted from {\tt RAPGAP}. 
The $b$-quark mass was set to $m_b = 4.75$ GeV and the renormalisation and factorisation
scales to $\mu = \sqrt{Q^2+4m_b^2}$. The {\tt CTEQ5F4} proton parton densities~\cite{Lai:1999wy}
were used. The NLO QCD prediction is $28^{+5.3}_{-3.5}$ pb, where the error was estimated 
by varying the scale $\mu$ by a factor of 2 and the mass $m_b$ between 4.5 and 5.0 GeV 
and adding the respective contributions in quadrature. A more detailed discussion of 
the QCD prediction and the related uncertainties for this process can be found in Ref.~\cite{Carli:2003cx}.
The measured total cross section is somewhat above but agrees with the NLO prediction
within the experimental and theoretical uncertainties.
The program {\tt CASCADE} \cite{Jung:2000hk}, based on the CCFM evolution equations~\cite{Ciafaloni:1987ur} 
and on gluon densities depending on the transverse parton momenta emitted along
the cascade, gives a prediction of 35 pb, which is in good agreement with the data.

\begin{figure}[!thb]
  \vspace*{7.0cm}
  \begin{center}
    \includegraphics{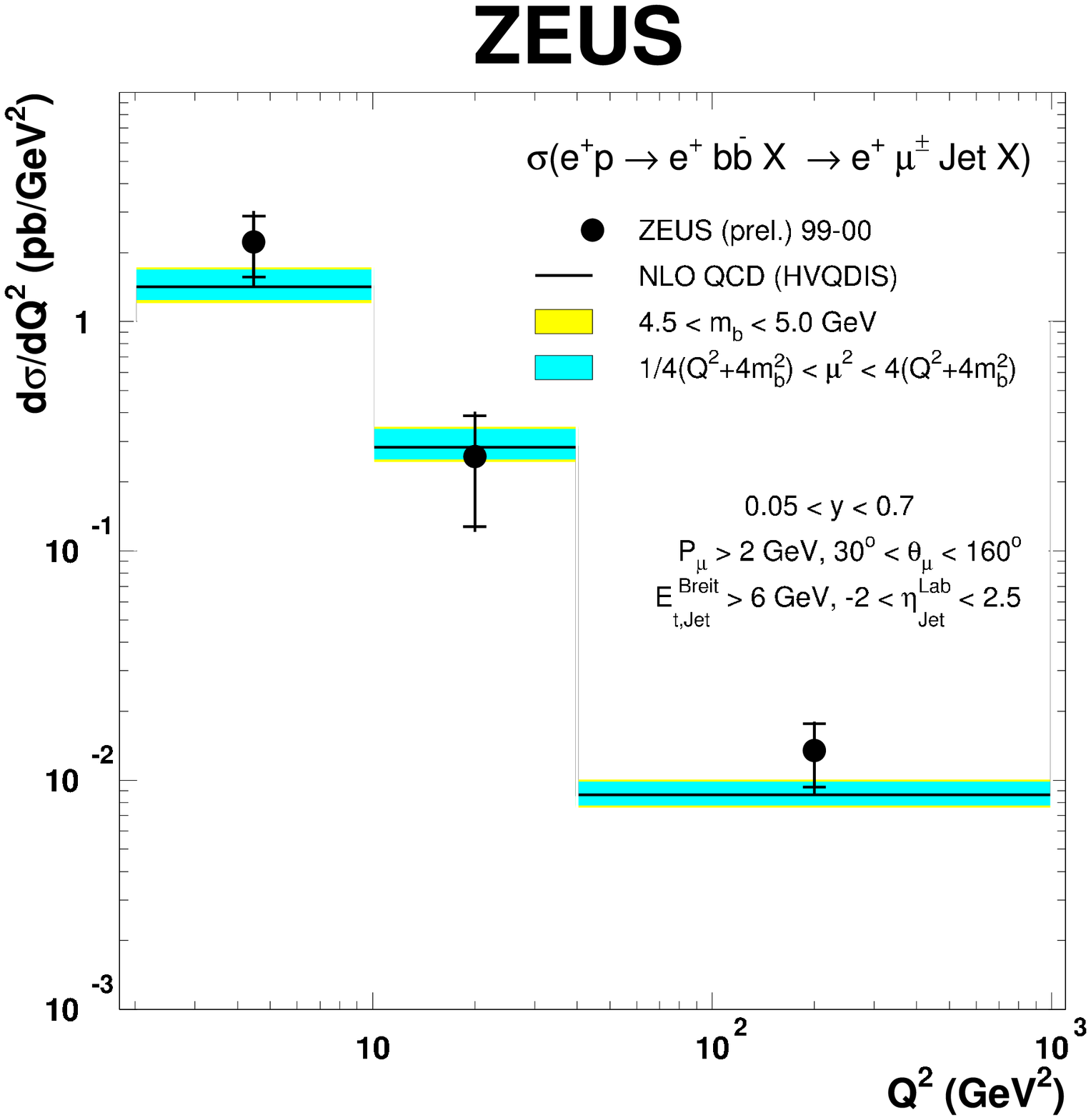}
    \hspace{1mm}
    \includegraphics{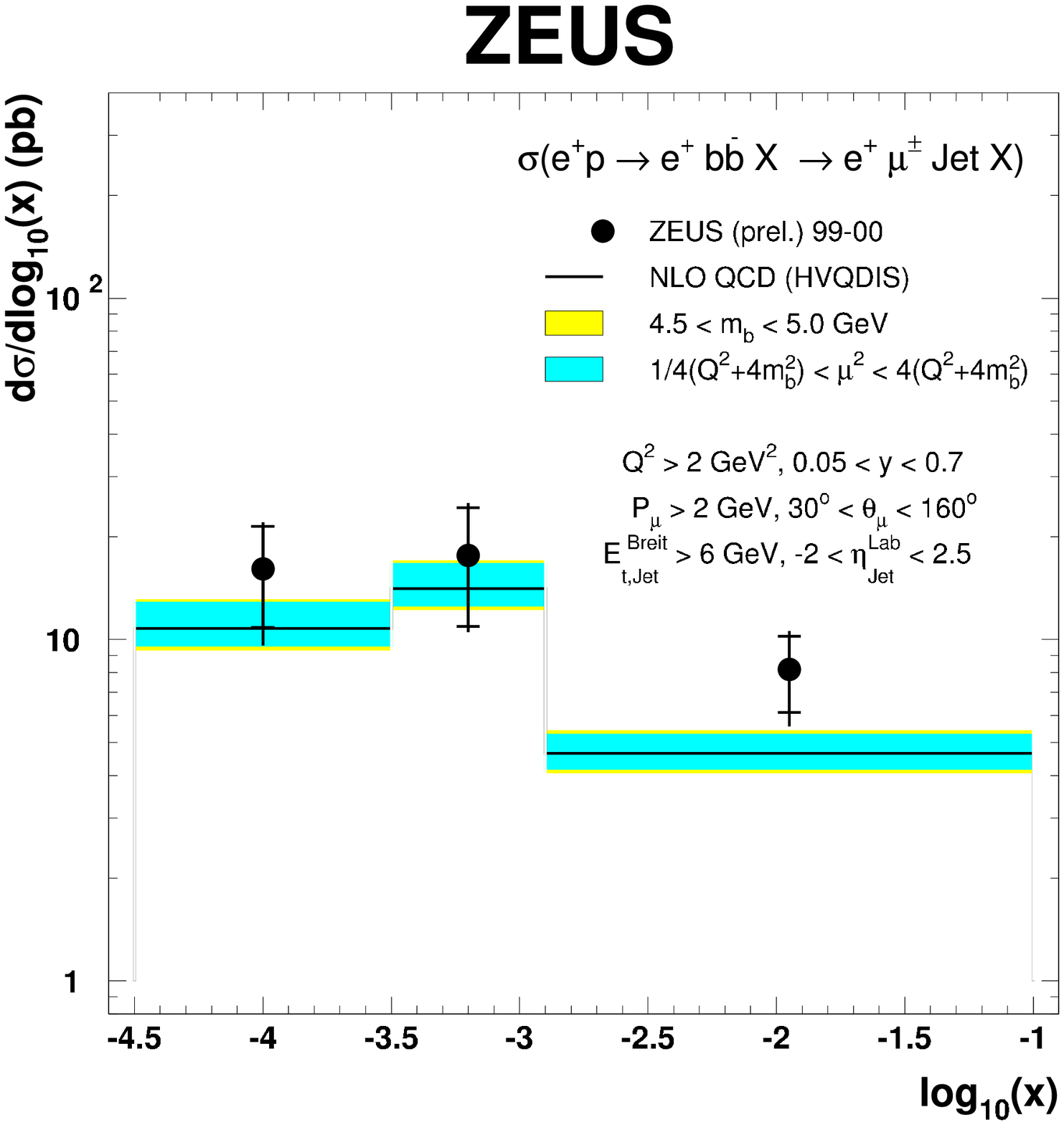}
    \vspace{-1.8cm}
    \caption{Differential beauty cross section as a function of $Q^2$ (left) and
      as a function of $x$ (right), compared to the NLO QCD calculations. The shaded
      bands show the uncertainty of the theoretical prediction ({\it see text}).}
    \label{FIG:DataNLO}
  \end{center}
\end{figure}

\begin{figure}[!thb]
  \vspace*{7.0cm}
  \begin{center}
    \includegraphics{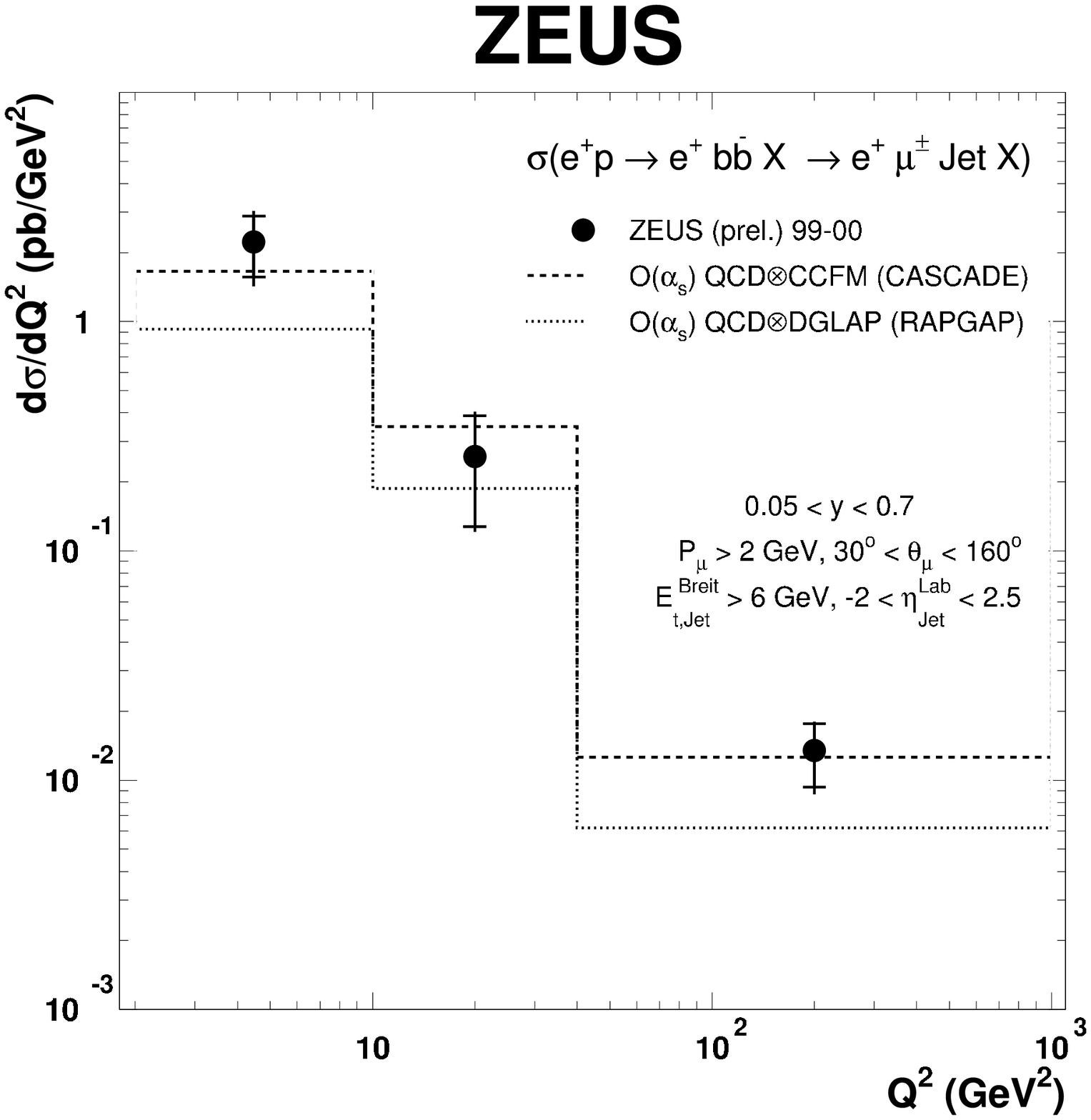}
    \hspace{1mm}
    \includegraphics{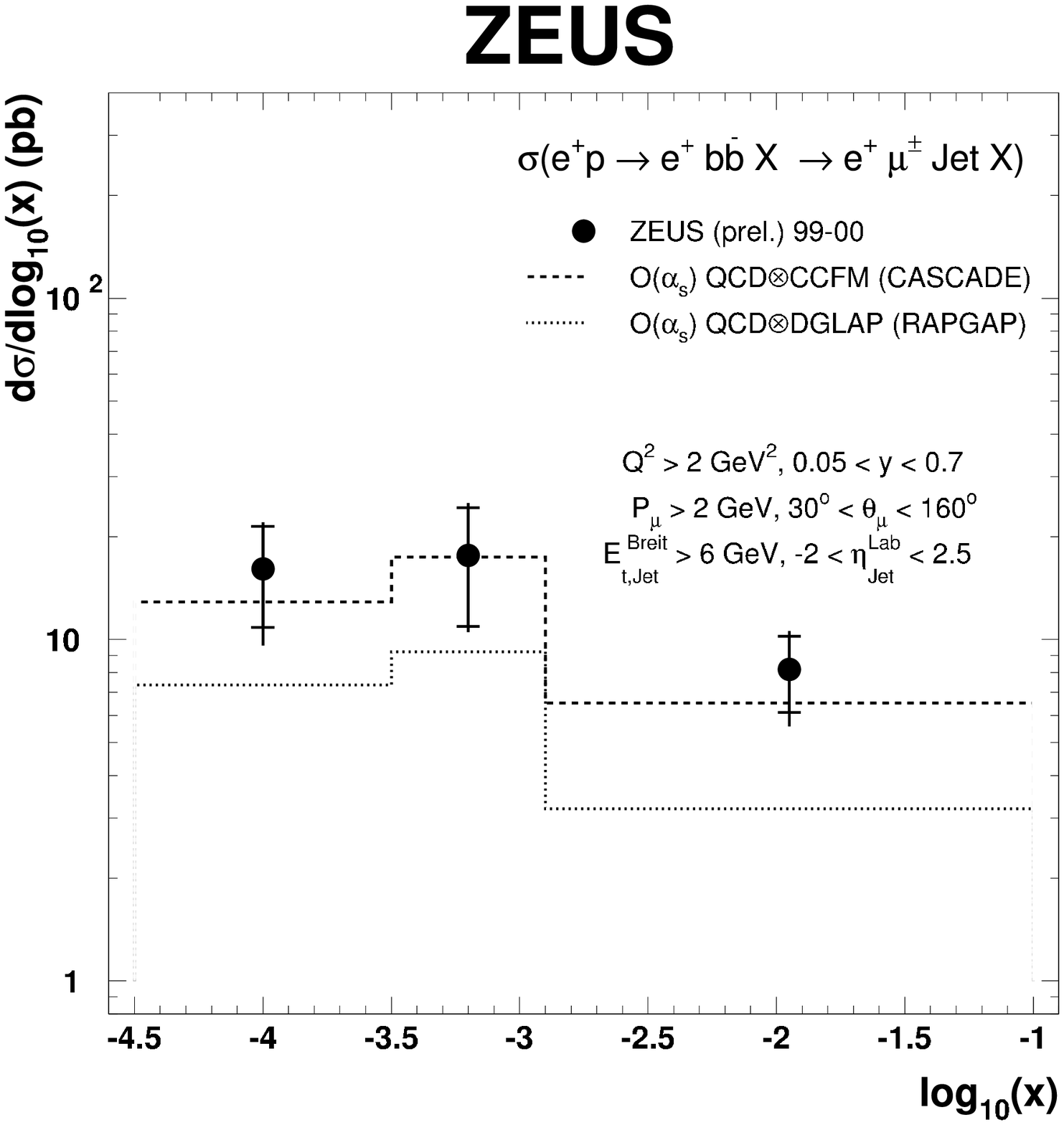}
    \vspace{-1.8cm}
    \caption{Differential beauty cross section as a function of $Q^2$ (left) and
      as a function of $x$ (right), compared to the LO QCD Monte Carlo programs.}
    \label{FIG:DataLO}
  \end{center}
\end{figure}

The differential cross sections were calculated in the same kinematic range
as the total cross section by repeating the fit of the $p_T^{rel}$ distribution in
each bin. Figure~\ref{FIG:DataNLO} shows the differential cross section as functions
of $Q^2$ and $x$ compared to the NLO. The NLO predictions agree reasonably well 
with the data.
Figure~\ref{FIG:DataLO} shows the same differential cross sections compared
with the LO Monte Carlo programs. The LO QCD prediction folded with the
DGLAP QCD \cite{Gribov:ri} evolution equations ({\tt RAPGAP}) underestimates the measured cross 
section. {\tt CASCADE} is in good agreement with the data.
\section{Conclusions}

The production of beauty quarks was measured in the process 
$e^+p \rightarrow e^+ + \mu + \mbox{jet} + X$ with the ZEUS detector at HERA.
The NLO QCD predictions agree with the measured total cross section and differential
cross sections as functions of $Q^2$ and the Bjorken variable $x$ within the
experimental and theoretical uncertainties. The {\tt CASCADE} Monte Carlo program 
gives a good description of the measured cross sections.

\section*{Acknowledgements} 
I would like to thank my collegues of the ZEUS collaboration
for their work and their suggestions to this talk.
%
%
%
%
%
%
%

\end{document}